\documentclass[a4paper,12pt]{article}

\usepackage{fancyhdr} 

\usepackage{comment}

\def\nbslash{\rlap{\hspace{0.02cm}/}{\bar n}}

\usepackage{multicol}
\usepackage{etoolbox}
\usepackage{cancel}
\usepackage[table]{xcolor}

\usepackage[font=small,labelfont=bf]{caption}
\DeclareCaptionFont{scriptsize}{\scriptsize}
\DeclareCaptionFont{tiny}{\tiny}
\usepackage{float}
\usepackage{amsmath, amssymb, setspace,cite,verbatim}
\usepackage{graphicx}
\usepackage{hyperref}
\usepackage{color}
\usepackage{wrapfig}

\usepackage{multicol}
\usepackage{etoolbox}
\usepackage{relsize}
\usepackage{multirow}

\usepackage{latexsym}
\usepackage{a4wide}
\usepackage{graphicx, subfigure}
\usepackage{cite}
\usepackage{tabularx}
\usepackage{array}
\usepackage{graphics}     
\usepackage{amsmath}
\usepackage{amssymb}
\usepackage{longtable} 
\usepackage{verbatim} 
\usepackage[table]{xcolor}
\usepackage{booktabs}
\usepackage{hyperref} 
\usepackage[utf8]{inputenc}
\usepackage{soul}

\usepackage{amsmath}
\usepackage{amssymb}

\usepackage{slashed}

\usepackage{pifont}

\definecolor{light-gray}{gray}{0.90}

\begin{document}

\hfill {\tt  MITP-25-080, DESY-25-182 }  

\def\thefootnote{\fnsymbol{footnote}}
 
\begin{center}

\vspace{3.cm}

{\Large\bf {Update of the nonlocal  sub-leading  ${\cal O}_1$  - ${\cal O}_7$  contribution to $\bar B \to X_s \gamma$ at LO}~\footnote{Addendum to  M.~B. and T.~H., Phys.~Rev.~D~\textbf{102} (2020), 114024 [arXiv:2006.00624 [hep-ph]].}}

\setlength{\textwidth}{11cm}
                    
\vspace{2.cm}
{\large\bf  
Michael Benzke$^{\,a}$\footnote{Email: michael.benzke@desy.de}, Maria Vittoria Garzelli$^{\,a}$\footnote{Email: maria.vittoria.garzelli@desy.de},
Tobias Hurth$^{b}$\footnote{Email: tobias.hurth@cern.ch},
}  
 
\vspace{1.cm}
{\em $^a$II. Institute for Theoretical Physics, University Hamburg\\ 
Luruper Chaussee 149, D-22761 Hamburg, Germany}\\[0.2cm]
{\em $^b$PRISMA+ Cluster of Excellence and  Institute for Physics (THEP)\\
Johannes Gutenberg University, D-55099 Mainz, Germany}\\[0.2cm] 

\end{center}

\renewcommand{\thefootnote}{\arabic{footnote}}
\setcounter{footnote}{0}

\vspace{1.cm}
\thispagestyle{empty}
\centerline{\bf ABSTRACT}
\vspace{0.5cm}
In all previous calculations of the  non-local sub-leading  contribution to the inclusive penguin decay $\bar B \to X_s \gamma$ due to the interference of the electroweak operators ${\cal O}_1^c$  - ${\cal O}_{7\gamma}$  the local Voloshin term was subtracted.  In view of the ongoing analysis  at order $\alpha_s$, we present a  calculation of the complete non-local  contribution which takes into account the high correlation between the uncertainties of the local Voloshin and the non-local term of the previous analyses. The new calculation
has a high impact on the range of the non-local contribution.

\clearpage

\section{Introduction}

The resolved contributions are nonlocal sub-leading contributions to the inclusive penguin modes, which are characterised by containing subprocesses in which the real or virtual photon couples to light partons instead of connecting directly to the effective electroweak vertex~\cite{Lee:2006wn}.
These contributions cannot be calculated in a local operator product expansion (OPE)  but must be treated using factorisation methods of soft collinear effective theory (SCET). Systematic analyses of these contributions have already been performed for both inclusive penguin modes, $\bar B \to X_s \gamma$ and $\bar B \to X_s \ell\ell$, to leading order in the strong coupling constant $\alpha_s$~\cite{Benzke:2010js,Benzke:2010tq,Hurth:2017xzf,Benzke:2017woq} 

Among the resolved contributions the sub-leading contribution due to the interference of the electroweak operators ${\cal O}^{c}_1$  - ${\cal O}_{7\gamma}$ is the numerically most important one. Additionally, the uncertainty due to this term is also among the largest within both inclusive penguin modes~\cite{Misiak:2015xwa,Huber:2015sra,Huber:2019iqf}.
One finds a large charm mass dependence and also a large scale ambiguity at leading order~\cite{Benzke:2020htm}. This calls for a  systematic calculation of $\alpha_s$ corrections and renormalisation group (RG) resummation~\cite{Benzke:2020htm}. For this task a factorisation formula for the sub-leading resolved corrections is needed which is valid to all orders in  $\alpha_s$. Concerning this point, another new input was given in Ref.~\cite{Hurth:2023paz} where a previous failure of factorisation in specific resolved contributions was healed by using new refactorisation techniques~\cite{Liu:2019oav,Liu:2020wbn,Beneke:2022obx}.

The factorisation theorem for the resolved ${\cal O}^{c}_1$  - ${\cal O}_{7\gamma}$ contribution has the schematic form 
\begin{equation}
\label{eq:LP-factorisation}
\frac{d\Gamma(\bar{B} \to X_s \gamma)}{dE_\gamma} \sim H \cdot J \otimes g \otimes \hat{J} \,,
\end{equation}
where $ H $ is the hard function encoding short-distance physics at the scale $ \mu_H \sim m_b $, $ J $  and $ \hat{J}$ are the jet functions describing (anti-)hard-collinear dynamics at the intermediate scale $ \mu_{J,\,\hat{J}} \sim \sqrt{m_b \Lambda_{\text{QCD}}} $, and $ g $ is the non-perturbative shape function capturing soft physics at the hadronic scale $ \mu_S \sim \Lambda_{\text{QCD}} $. The symbol $ \otimes $ denotes a convolution in light-cone momentum.

All functions can be calculated perturbatively except the soft function. 
In our present case at the sub-leading power level the shape function ${g}_{17}(\omega,\omega_1,\mu)$ is given by the non-local heavy-quark effective theory (HQET) matrix element in Ref.~\cite{Benzke:2010js}:
\begin{eqnarray}\label{g17def}
   g_{17}(\omega,\omega_1,\mu) 
   &=& \int\frac{dr}{2\pi}\,e^{-i\omega_1 r}\!
    \int\frac{dt}{2\pi}\,e^{-i\omega t} \\
   &&\times \frac{\langle\bar B| \big(\bar h S_n\big)(tn)\,
    \nbslash (1+\gamma_5) \big(S_n^\dagger S_{\bar n}\big)(0)\,
    i\gamma_\alpha^\perp\bar n_\beta\,
    \big(S_{\bar n}^\dagger\,g G_s^{\alpha\beta} S_{\bar n} 
    \big)(r\bar n)\,
    \big(S_{\bar n}^\dagger h\big)(0) |\bar B\rangle}{2M_B} \,,
    \nonumber
\end{eqnarray}
 where $n$ and $\bar n$ are the light-cone vectors, $h$ the heavy quark field and $G$ the gluon field tensor. $S_{n, \bar n}$ denote soft Wilson lines in both light cone directions. They assure the gauge invariance of the expression.
The shape function depends on two variables and is non-local in both light-cone directions. This significantly complicates the renormalisation of this soft function. The renormalisation and also the RG running of the soft function has been worked out in Ref.~\cite{Bartocci:2024bbf,Bartocci:2025lbl}.  This is the first step towards the next-to-leading (NLO) radiative corrections of the resolved contribution due to the interference of ${\cal O}^{c}_1$~-~${\cal O}_{7\gamma}$.

A  systematic {method} to estimate the value of the convolution integral of the perturbative jet functions and the non-perturbative shape function  $h_{17}$ was first introduced to this problem in Ref.~\cite{Gunawardana:2019gep}: one derives all general properties of the shape function and then uses a complete set of basis functions, for example the Hermite functions multiplied by a Gaussian function, in order to make a systematic analysis of all possible model functions fulfilling the known properties of the shape function. This kind of systematic approach to shape functions was already used in several previous analyses~\cite{Ligeti:2008ac,Lee:2008xc,Bernlochner:2020jlt}.

Concretely, one shows PT invariance of the soft function which implies that the matrix element is real~\cite{Benzke:2010js} . Moreover, one can calculate  moments of this HQET matrix element: while the zeroth moment was already known in the first analyses of the resolved contributions, the second moment was more recently derived using HQET techniques; moreover, rough dimensional estimates of the higher-order moments were proposed~\cite{Gunawardana:2017zix,Gunawardana:2019gep}. In addition one can also naturally assume that the support properties and the values of the non-perturbative shape function are within the hadronic range. 
Obviously this systematic approach allows to avoid any prejudice regarding the unknown functional form of the shape functions and, thus, leads to a valid estimate of the resolved contribution. Any additional assumption calls for a clear justification.

The calculation is done for large photon energies near the kinematical endpoint, \mbox{$m_b - 2E_\gamma = \mathcal{O}(\Lambda_{\rm QCD}) \ll m_b$}. In this region, the hadronic state $X_s$ has a large energy of the order $\mathcal{O}(m_b)$, but small invariant mass of the order $\mathcal{O}(m_b \Lambda_{\rm QCD}) \ll m_b^2$, which is still a perturbative scale. A local OPE fails to describe this jet-like configuration, but the latter can be analysed with factorisation methods from SCET.  Moreover, the effects of the resolved contributions still have to be described in terms of non-local operator matrix elements when the photon cut is moved outside the endpoint region~\cite{Benzke:2010js,Hurth:2017xzf,Benzke:2017woq}. In addition, the support properties of the shape functions imply that the resolved contributions -- except the ${\cal O}_8 - {\cal O}_8$ one --  are almost cut independent. Therefore  all the resolved contributions represent an irreducible uncertainty.

\section{Strategy of  the calculation}
In Ref.~\cite{Benzke:2020htm} we presented a systematic analysis of the dominant resolved contribution due to  the interference ${\cal O}^{c}_{1}$ - ${\cal O}_{7\gamma}$ to the inclusive penguin decays $\bar B \to X_{s} \gamma$ following {the method} proposed in Ref.~\cite{Gunawardana:2019gep} .

Using the original notation of Ref.~\cite{Benzke:2010js} one can write {the resolved contribution due to  the interference   ${\cal O}^{c}_{1}$ - ${\cal O}_{7\gamma}$} normalised to the perturbative leading order (LO) result as 
\begin{equation}\label{relative uncertainty}
  {\cal F}_{\rm b \to s \gamma}^{17} = \frac{C_1(\mu)\, C_{7\gamma}(\mu)}{C_{7\gamma}(\mu_{\rm \mbox{{\tiny OPE}}})^2}\, \frac{\Lambda_{17}(m_c^2/m_b,\mu)}{m_b}\,, 
\end{equation}
where $\mu_{\rm \mbox{{\tiny OPE}}}$ denotes the hard scale within the local OPE and $\mu$ the scale within the resolved contribution.  As in the original analysis in Ref.~\cite{Benzke:2010js}, the perturbative decay rate at the hard scale at LO accuracy is used as normalisation, meaning  that also the Wilson coefficient is taken at the hard scale $\mu_{\rm \mbox{{\tiny OPE}}}$. 
Since no $\alpha_s$ corrections or any RG improvements are considered yet in the calculation of the resolved power corrections, the scale choice is ambiguous. One first fixes the Wilson coefficients in the resolved contribution at the hard scale. Then one varies the scale of the Wilson coefficients in the resolved contributions between the hard and the hard-collinear scale to make the scale ambiguity of the resolved contributions manifest.

 At subleading power one finds~\cite{Benzke:2010js}:
\begin{equation}\label{Lambda17A}
  \Lambda_{17}\Big(\frac{m_c^2}{m_b},\mu\Big)
  = e_c\,\mbox{Re} \int_{-\infty}^\infty \frac{d\omega_1}{\omega_1} 
  \left[ 1 - F\!\left( \frac{m_c^2-i\varepsilon}{m_b\,\omega_1} \right)
  + \frac{m_b\,\omega_1}{12m_c^2} \right] h_{17}(\omega_1,\mu)\,,
\end{equation}
with the penguin function $F(x) = 4\, x\, {\rm arctan}^2(1/\sqrt{4x-1})$. The shape function $h_{17}$ is connected to $g_{17}$ defined in Eq.~\ref{g17def}  in the following way:
\begin{equation}
   h_{17}(\omega_1,\mu) 
   = \int_{- \infty}^{\bar\Lambda}\!d\omega\,
    g_{17}(\omega,\omega_1,\mu) \,,\\
\end{equation}
where $\omega_1$ corresponds to the soft gluon momentum and $\bar \Lambda=M_B-m_b$.  

In Eq.~(\ref{Lambda17A}) the third term in the brackets corresponds to the local Voloshin term~\cite{Voloshin:1996gw} (multiplied by $-1$), which is traditionally subtracted from the $Q_1^c - Q_{7\gamma}$ resolved contribution. This term is proportional to the zeroth moment of the shape function $g_{17}$ which can be expressed by the well-known HQET parameter $\lambda_2$:~\footnote{This local Voloshin term can be derived from the complete resolved contribution ${\cal O}^{c}_{1} - {\cal O}_{7\gamma}$  by neglecting the shape function effects and under the assumption that the charm quark mass is treated as heavy. It does not account for the complete resolved contribution.}
\begin{equation}\label{g17norm}
   \int_{-\infty}^{\bar\Lambda}\!d\omega 
   \int_{-\infty}^\infty\!d\omega_1\,g_{17}(\omega,\omega_1,\mu) 
   = \frac{\langle\bar B|\,\bar h\,\nbslash\,i\gamma_\alpha^\perp
           \bar n_\beta\,g G_s^{\alpha\beta}\,h\,|\bar B\rangle}{2M_B} 
   = 2\lambda_2 \,.
\end{equation}
Integrating the third term und taking into account the minus sign we get 
\begin{equation}
 \Lambda_{17}^{\rm Voloshin} = (-1) \frac{m_b \lambda_2}{9 m_c^2} \,.
 \end{equation}
This formula translates via Eq.~(\ref{relative uncertainty}) into  
\begin{equation} \label{Voloshin}
{\cal F}_{17}^{\rm Voloshin} = - \frac{C_1(\mu)\,C_{7\gamma}(\mu)\, \lambda_2}{C_{7\gamma}(\mu_{\rm OPE})^2\, 9\, m_c^2} = +3.3\%\,.
\end{equation}
This normalised central value of the Voloshin term was evaluated using the LO Wilson coefficients and the hard OPE scale in the resolved part.

From now on we denote the quantity in Eq.~(\ref{Lambda17A}) as shape function 
term $\Lambda_{17}^{\rm Shape}$.
We can then write the complete resolved contribution as 
\begin{equation} \label{Separation}
\Lambda_{17}^{\rm Complete} = \Lambda_{17}^{\rm Shape}  + \Lambda_{17}^{\rm Voloshin}  = e_c\,\mbox{Re} \int_{-\infty}^\infty \frac{d\omega_1}{\omega_1} 
  \left[ 1 - F\!\left( \frac{m_c^2-i\varepsilon}{m_b\,\omega_1} \right) \right] h_{17}(\omega_1,\mu)\,.
\end{equation}
The separate evaluation of the local Voloshin term $\Lambda_{17}^{\rm Voloshin}$ within the ${\cal O}_1$  - ${\cal O}_7$ resolved contribution was presented a long time ago~\cite{Voloshin:1996gw,Ligeti:1997tc,Grant:1997ec,Buchalla:1997ky}. 
The latest calculation of the shape function term  $\Lambda_{17}^{\rm Shape}$  was done in Ref.~\cite{Benzke:2020htm} and leads  to -- including uncertainties due to all parameters~\footnote{The authors of Ref.~\cite{Gunawardana:2019gep} found much smaller values. As discussed in detail in Ref.~\cite{Benzke:2020htm}, the main reasons for this difference is the much smaller variation of the charm mass and the non-inclusion of an uncertainty due to $1/m_b^2$ corrections in Ref.~\cite{Gunawardana:2019gep}.} --
%
\begin{equation} \label{resultshape} 
-61  \,\text{MeV} \leq \Lambda_{17}^{\rm Shape}  \leq +5\,\text{MeV}, \hspace{1cm} 
\end{equation} 
which translates via Eq.~(\ref{relative uncertainty}) into
\begin{equation}
 {\cal F}^{\rm Shape}_{17} \in  [-0.4\%,\,4.7\%]\,,       \label{resultbsgamma1}
 \end{equation}
where we used again the LO Wilson coefficients and $m_b=4.58\text{ GeV}$. 

Adding the central value of the Voloshin term without uncertainties (see Eq.~(\ref{Voloshin})) we get
\begin{equation}
 {\cal F}^{\rm Shape}_{17}  +  {\cal F}^{\rm Voloshin}_{17} \in  [2.9\%,\,8\%]\,.       \label{resultbsgamma2}
 \end{equation}

In addition, a substantial scale ambiguity must be considered. At LO, the scale of the hard functions, i.e. of the Wilson coefficients in the resolved contribution, is not fixed. We make this scale ambiguity manifest by shifting the scale of the Wilson coefficients from the hard scale to the hard-collinear scale. This results in an increase of the range by more than $40\%$ to $[4.2\%,11.7\%]$.  The overall range including the scale ambiguity is then 
\begin{equation}
 {\cal F}^{\rm Shape}_{17}  +  {\cal F}^{\rm Voloshin}_{17} \in  [2.9\%,\,11.7\%]\, \hspace{1cm}  ({\rm  scale\, ambiguity\, included)}  \,.
 \label{resultbsgamma3}
 \end{equation}
{We emphasise that the uncertainties of the Voloshin term due to parameter variation are {not} taken into account yet.}
Furthermore, Eq.~(\ref{Separation}) shows that the separation of these two terms is unnatural from the SCET point of view  and it is clear that an error analysis of the complete ${\cal O}_1^c$  - ${\cal O}_{7\gamma}$ resolved contribution has to be redone  because the uncertainties of the two terms are highly correlated.

\section{Numerical analysis}
We calculate now the complete resolved ${\cal O}^{c}_{1} - {\cal O}_{7\gamma}$ contribution without separating this  contribution into two pieces.  
As in our previous analysis in Ref.~\cite{Benzke:2020htm}, we follow the systematic method proposed in Ref.~\cite{Gunawardana:2019gep} and use a complete set of basis functions, namely the Hermite polynomials multiplied by a Gaussian:
\begin{equation}\label{Hermite} 
h_{17}(\omega_1)=\sum_n a_{2n} H_{2n}\left(\frac{\omega_1}{\sqrt{2}\sigma}\right)e^{-\frac{\omega_1^2}{2\sigma^2}}. 
\end{equation}
Here only even polynomials have to  be chosen due to the fact that our shape function $h_{17}$ is even. We also allow for higher suppression in the Gaussian because such functions can not be expressed by the basis of functions given in Eq.~(\ref{Hermite})  and, thus, have to be included within a systematic analysis. 
This approach avoids any prejudices regarding the unknown functional form of the shape function, but only uses well established properties like the zeroth and the second moment {which were derived in                               Refs.~\cite{Gunawardana:2017zix,Gunawardana:2019gep}}, and also the fourth and the sixth moment which are based on dimensional arguments.

All input parameters are varied on our grid  {within their uncertainties, as done in the previous analyses~\cite{ Gunawardana:2019gep,Benzke:2020htm}, because we cannot assume that the uncertainties are in general Gaussian in the considered ranges and therefore we do not add them in quadrature.} The charm quark mass is varied within the range $1.14\,\text{GeV} \leq m_c \leq 1.23\,\text{GeV}$ in 10 steps, the bottom quark mass within the range $4.55\,\text{GeV} \leq m_b \leq 4.61\,\text{GeV}$ in 3 steps. The zeroth moment $m_0 = 2 \lambda_2$  is varied from  $0.197\,{\rm GeV^2}$ to $0.277\,{\rm GeV^2}$ in 8 steps and the second moment $m_2$ from $0.03\,{\rm GeV^4}$ to $0.27\,{\rm GeV^4}$ in 12 steps; the fourth and the sixth moment within $-0.3\,{\rm GeV^6} \leq m_4  \leq    0.3\,{\rm GeV^6}$ and within  $-0.3\,{\rm GeV^8} \leq m_6 \leq 0.3\,{\rm GeV^8}$, respectively, in 30 steps. The parameter $\sigma$ of the model functions in Eq.~(\ref{Hermite}) is varied from $0\, \text{GeV}$ to $1\, \text{GeV}$ via 20 steps. For details on the schemes adopted for the charm quark mass and the bottom quark mass we refer the reader to Ref~\cite{Benzke:2020htm}.

{For a polynomial of degree $6$ with an  ${\rm exp}(-x^4)$ suppression, like in the calculation of the shape function term in Ref.~\cite{Benzke:2020htm}, we find the following range for the complete resolved contribution. }
\begin{equation} \label{finalresult} 
- 115 \,\text{MeV} \leq \Lambda_{17}^{\rm Complete}  \leq - 34\,\text{MeV}\,,
\end{equation} 
which translates via Eq.~(\ref{relative uncertainty}) into
\begin{equation}
 {\cal F}^{\rm Complete}_{17} \in  [2.6\%, 8.9\,\%]\,,       \label{resultbsgammax1}
 \end{equation}
which can be compared to Eq.~(\ref{resultbsgamma2}) where the uncertainty of the Voloshin term was not included.
{The range increases by  $22.5\%$.}

Polynomials of degree $8$ or higher suppression factors like ${\rm exp}(-x^6)$ do not lead to more extreme values.   

It turns out that the grid parameters which lead to the two extreme values for the complete resolved contribution are the same as in the case of the extreme values for the shape function term only, found in our previous analysis in Ref.~\cite{Benzke:2020htm}. The latter result, quoted in Eq.~(\ref{resultshape}), has been also confirmed by the new calculation. 

The grid parameters for the largest value are $m_c =  1.14\, \text{GeV}$, $m_b = 4.61\, \text{GeV}$, $\sigma = 0.750\, \text{MeV}$, $m_0 = 0.197\,\text{GeV}^2$, $m_2 = 0.030\,\text{GeV}^4$, $m_4 = - 0.100\,\text{GeV}^6$, $m_6 = - 0.180\,\text{GeV}^8$  and for the smallest  the parameters are  $m_c =  1.14\, \text{GeV}$, $m_b = 4.61$\, \text{GeV}, $\sigma = 0.700\, \text{MeV}$, $m_0 = 0.277\,\text{GeV}^2$, $m_2 = 0.270\,\text{GeV}^4$, $m_4 = 0.280\,\text{GeV}^6$, $m_6 = 0.300\,\text{GeV}^8$.

A specific pattern has also been seen in our previous analysis of just the shape function term. Both extreme values are found for the smallest $m_c$ value and for the largest $m_b$ value. This can be understood by the fact that these values allow for a maximum overlap of the shape function with the jet function. The fact that the grid parameters corresponding to the extreme values of the complete resolved term and of just the shape function term are the same implies that the increase of the range can solely traced back to the Voloshin term including its uncertainties.

The large scale ambiguity of the resolved contribution at LO can be made manifest again by shifting the scale of the Wilson coefficients from the hard to the hard-collinear scale. This leads to the range $[3.8\%,13\%]$.  Thus,  the overall range including the scale ambiguity is then 
\begin{equation}
 {\cal F}^{\rm Complete}_{17}  \in  [2.6 \%,\, 13\%]   \hspace{1cm}   ({\rm  scale\, ambiguity\, included)}  \,.   \label{finalresultA} 
 \end{equation}
 This represents our final LO result for the complete resolved   ${\cal O}_1^c$  - ${\cal O}_{7 \gamma}$  contribution.
{We note that the range does not correspond to a Gaussian uncertainty. The estimate  should be considered as the range within which we expect the actual values of the resolved contribution  to lie, without making a statement about the most likely values within this range  --- as noted already in Ref~\cite{Benzke:2010js}.}

\section{Conclusions}

We presented a calculation of the complete nonlocal resolved contribution due to the interference of the electroweak operators ${\cal O}_1^c$  - ${\cal O}_{7 \gamma}$ at leading order (LO).
This joint calculation of the shape function term and the local Voloshin term is mandatory due to the fact that the uncertainties are highly correlated. The new calculation  leads to a significant increase of the range of the resolved contribution which does not correspond to a Gaussian error.  {We find ${\cal F}^{\rm Complete}_{17}  \in  [2.6 \%,\, 8.9\%] $.  This represents one of the largest uncertainties of the inclusive radiative penguin decay $\bar B \to X_s \gamma$. The large error can be traced back to  a large charm mass uncertainty in the LO result.   If we make the scale ambiguity manifest  we get an even larger range, namely  ${\cal F}^{\rm Complete}_{17}  \in  [2.6 \%,\,13\%] $}. The ongoing $\alpha_s$ calculation~\cite{Bartocci:2024bbf,Bartocci:2025lbl} will establish a well-defined scale dependence diminishing the large scale ambiguity and may reduce the large charm mass dependence.

\section*{Acknowledgements}
This work was supported by the Cluster of Excellence ``Precision Physics, Fundamental Interactions, and Structure of Matter" (PRISMA$^+$ EXC 2118/1) funded by the German Research Foundation (DFG) within the German Excellence Strategy (Project ID 39083149) {and by the FOR2926 Resarch Unit (Project ID 40824754) also funded by DFG}. TH thanks  the II. Institute for Theoretical Physics at University of Hamburg as well as the CERN theory group for hospitality during his regular visits to Hamburg and to CERN where part of this work was done.

\end{document}